\documentclass[a4paper]{article}

\usepackage{atmohead2013}
\usepackage[english]{babel}

\title{Remote control and telescope auto-alignment system for multiangle LIDAR under development at CEILAP, Argentina}

\shorttitle{AtmoHEAD 2013 Template}

\authors{
Juan Pallotta$^{1}$,
Pablo Ristori$^{1}$,
Lidia Otero$^{1}$,
Fernando Chouza$^{1}$,
D'El\'{i}a Ra\'{u}l$^{1}$,
Francisco Gonzalez$^{1}$,
Alberto Etchegoyen$^{2}$,
Eduardo Quel$^{1}$
for the CTA consortium.
}

\afiliations{
$^1$ CEILAP-UNIDEF-(MINDEF-CONICET), UMI-IFAECI-CNRS (3351) - Buenos Aires, Argentina. \\
$^2$  ITeDA (CNEA – CONICET - UNSAM) - Buenos Aires, Argentina. \\
}

\email{juanpallotta@citedef.gob.ar}

\abstract{At CEILAP (CITEDEF-CONICET), a multiangle LIDAR is under development to monitor aerosol extinction coefficients in the frame of the CTA (Cherenkov Telescope Array) Project. This is an initiative to build the next generation of ground-based instruments to collect very high energy gamma-ray radiation (\textgreater 10 GeV). The atmospheric conditions are very important for CTA observations, and LIDARs play an important role in the measurement of the aerosol optical depth at any direction. The LIDAR being developed at CEILAP was conceived to operate in harsh environmental conditions during the shifts, and these working conditions may produce misalignments. To minimize these effects, the telescopes comprising the reception  unit are controlled by a self-alignment system. This paper describes the self-alignment method and hardware automation.}

\keywords{multiangle LIDAR, Raman, CTA observatory, aerosols}
\begin{document}
\maketitle

%Begin a section.
\section{Introduction}

The Cherenkov Telescope Array Consortium (CTA) \cite{bib:cta_web} contemplates the design, construction and the operation of two observatories for the detection of gamma ray produced by extraterrestrial sources at energies ranging between $10^{10}$ eV and $10^{14}$ eV. These observatories will be deployed at each hemisphere for full sky-map coverage. Each Observatory will consist of a telescope array sensitive to the atmospheric generated Cherenkov radiation that will improve the performance of the actual detectors. The goals proposed for CTA will be attained using an array of multiple telescopes distributed over a surface of 1 and 10 km$^2$ in the northern and southern hemispheres, respectively located at sites with excellent optical and atmospheric conditions at a height of 1500  to 3800 m above the sea level. The comprehension of the atmospheric conditions during the measurements is extremely important for the CTA Observatory, and multiangle LIDARs plays a major role in monitoring of sky conditions, by both detecting the overall cloud coverage and measuring the atmospheric opacity due to aerosol and clouds. The multiangle Raman LIDAR being built at CEILAP uses six 40 cm f/2.5 Newtonian telescopes as a reception system. The additional feature added is an auto-alignment mirror system to follow the line of sight of the observation during the acquisition period. The system was designed to operate in harsh environmental conditions, as it is completely exposed to the weather when its shelter-dome is fully opened to provide 360 degrees observations. In this work, a description of the progress made on the development of the LIDAR automation system is provided.

\subsection{LIDAR Hardware}

The main features of the multiangle Raman LIDAR being built at CEILAP, are:

\begin{itemize}

\item {\bf Emission system:} Continuun Inlite II-50. It is a Nd:Yag laser, that generates laser pulses at 355, 532 and 1064 nm with a repetition rate of 50 Hz and a pulse energy of 125 mJ @ 1064  nm. Its divergence is assured to be less than 750 $\mu$rad.
\item {\bf Reception system:} Made by six Newtonian telescopes of 40 cm diameter and 1 m focal length each. An optical fiber of 1 mm diameter is placed at its focal plane, producing a 1 mrad field-of-view (FOV).
\item {\bf Spectrometric box:} A multiwavelength spectrometer separates the backscattered elastics and Raman lines. Raman lines used are: 387 nm and 607 nm (nitrogen backscatter from 355 nm and 532 nm respectively) and 408 (water vapour from 355 nm).
\end{itemize}

Additionally, this LIDAR has special requirements regarding its operation:
\begin{itemize}
\item It has to be operated remotely and the operator may not have an a priori knowledge of LIDAR techniques. 
\item Telescopes, mechanics and electronics, will be exposed during night time to environmental conditions (wind bursts, temperature spans, etc.), which could produce mirror misalignments.
\end{itemize}

\begin{figure}[h!]
  \centering
  \includegraphics[width=0.4\textwidth]{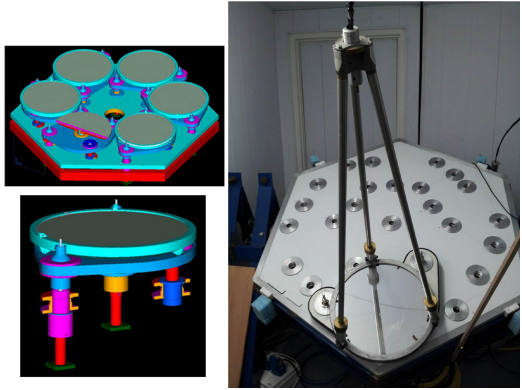}
  \caption{Telescopes mounting system layout and a picture of the system during test state.}
  \label{disenoFotoTelescopio_fig}
 \end{figure}

To minimize the signal loss due to misalignments, each of the six mirrors are mounted on a steerable frame, equipped with two stepper motors for tilting over two orthogonal axes (Figure \ref{disenoFotoTelescopio_fig}). These movements are handled by a microcontroller that communicates with the LIDAR PC through a WiFi connection (a brief description of this system is provided in section \ref{LIDARCommunication_section}).

\begin{figure}[h!]
  \centering
  \includegraphics[width=0.4\textwidth]{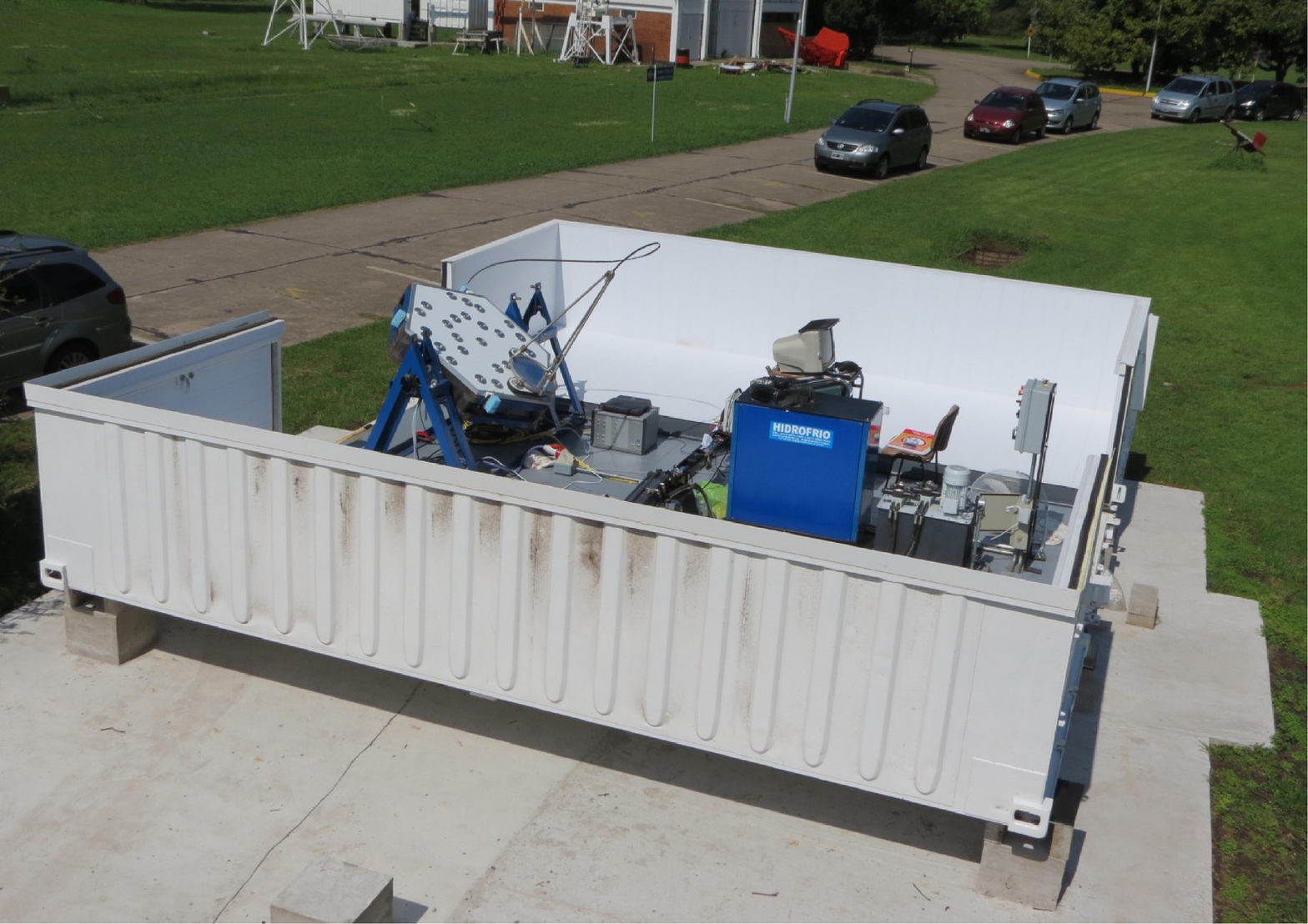}
  \caption{Full view of the prototype multiangle LIDAR in its shelter.}
  \label{LIDAR_contenedor_fig}
 \end{figure}

Since 2012, a shelter-dome was acquired to host the system. Based on the CLUE shelter concept, we have built this unit using a standard 20 ft shelter modified completely as shown in the Figure \ref{LIDAR_contenedor_fig}. CLUE shelters are also being used by other LIDAR groups in the CTA collaboration \cite{bib:LIDARDoroAtmoHead}.

To open/close the shelter, hydraulic cylinders were installed and can be controlled from its control panel located at the shelter, and remotely via the control software being developed at CEILAP.
These are the main reasons to encourage the development of a fully automatic alignment system to keep the telescopes aligned during the acquisition period.

\subsection{LIDAR Comunication} \label{LIDARCommunication_section}

\begin{figure}[h!]
  \centering
  \includegraphics[width=0.4\textwidth]{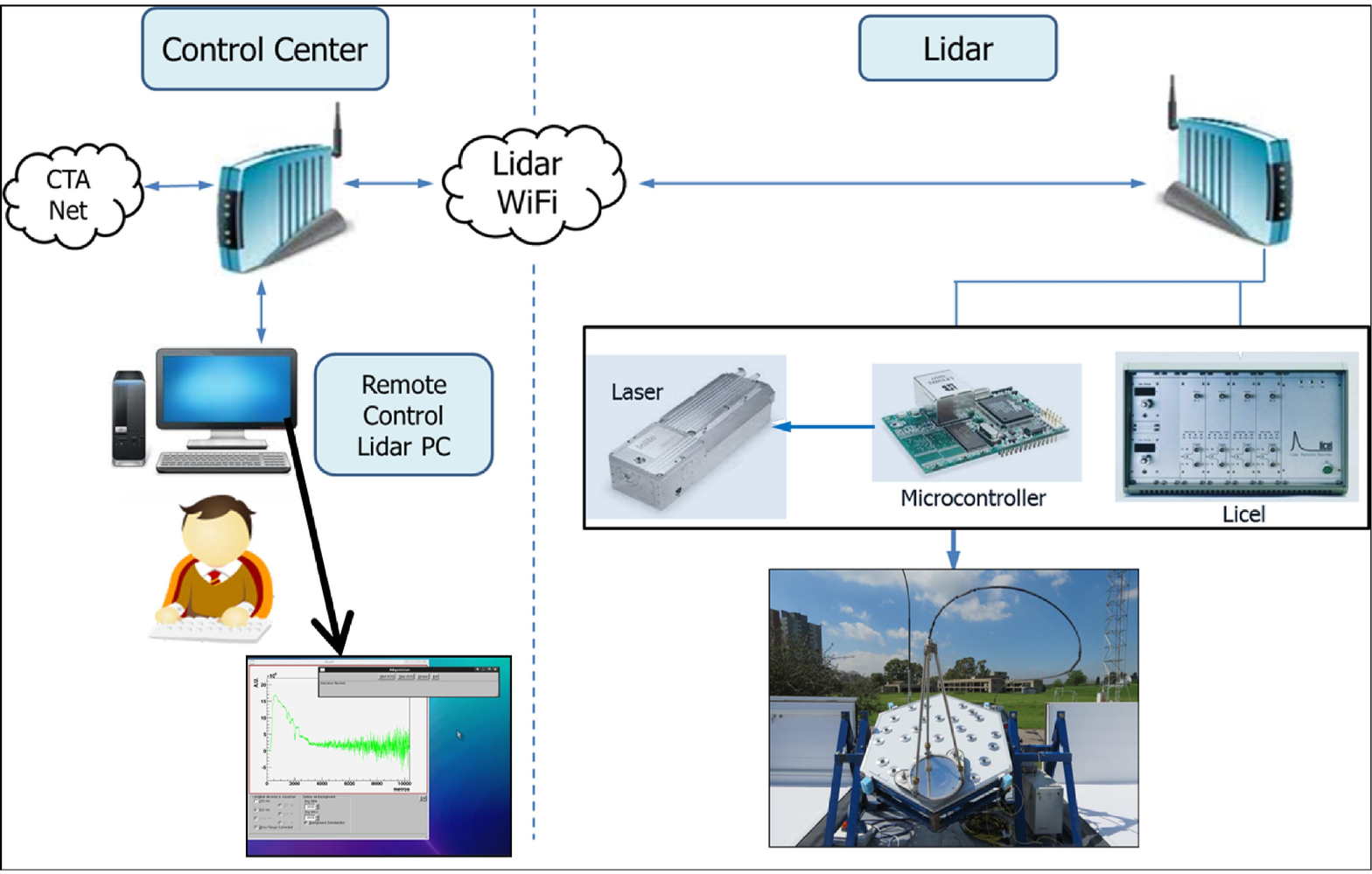}
  \caption{Communication layout between control PC and multiangle LIDAR.}
  \label{conexionWiFi_fig}
 \end{figure}

The LIDAR system under development has to be operated remotely from the control center, with a high degree of automation, as the local observing crew may have no advanced LIDAR training. The PC control LIDAR communicates with the LIDAR system via WiFi link, with two routers with WDS features paired, creating a local LIDAR network under the TCP/IP protocol (see Figure \ref{conexionWiFi_fig} for a schematic layout of the system).

At the link endpoints, several processes communicate with each other to send/receive control and monitoring messages, as will be briefly described on the next section. 

\subsection{LIDAR Software}

PC LIDAR control works under the Linux operating system and all the software was developed using C/C++. A socket-based IPC (Inter Process Communication) was programed to communicate with the microcontroller at the LIDAR shelter.
To increase their efficiency, each process is totally independent, and communicates to the others via control messages. The graphical user interfaces were developed using a set of ROOT libraries \cite{bib:root_web}. A brief description of each process is described below:\\
{\bf adq:} It is the main process, which controls the acquisition timing, communicates with the laser,  triggers the Licel acquisition system \cite{bib:licel_web}, sends the acquired new file to the {\bf plot} process, and, if necessary, to the alignment process.\\
{\bf plot:} Waits for messages from the {\bf adq} process, and plot the signals on the display.\\
{\bf alignment:} This process receives the path to the acquired file from {\bf adq} and processes this signal to obtain the alignment parameters to determine the telescope position. A brief description of this algorithm is described in section \ref{AutoAlignment_section}.\\

As can be seen in next figure, both {\bf adq} and {\bf plot} process are totally independent. {\bf alignment} does not require a graphical user interface.

 \begin{figure}[h!]
  \centering
  \includegraphics[width=0.4\textwidth]{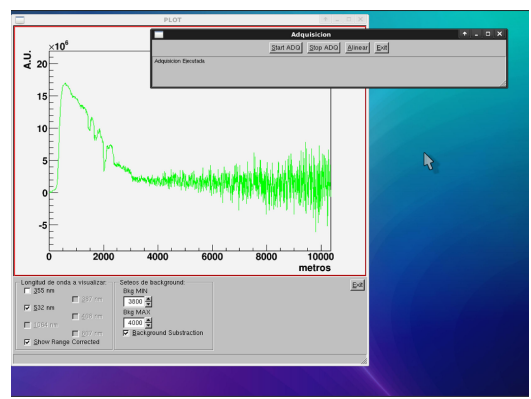}
  \caption{A print-screen of the LIDAR software. A 532 nm range corrected LIDAR signal is shown.}
  \label{soft_fig}
 \end{figure}

\subsection{Microcontroller-Controlled Telescopes}

This is a cooperative procedure between the {\bf adq} and the {\bf alignment} process, both running on the LIDAR PC, and the firmware inside the Rabbit microcontroller. The tilt angle of the telescopes is driven by a set of stepper motors, handled by a RCM2200 Rabbit System microcontroller \cite{bib:rabbit_web}. This is a Z80 family-based high-performance 8 bit microcontroller. It has a built-in Ethernet interface with an integrated TCP/IP stack, making it a good choice for interconnectivity. This interface is used to link the microcontroller with the LIDAR PC. The instruction set is based on the original Z80 microprocessor, with some additional instructions.
The aims of the Rabbit microcontroller algorithm are to decode the information received from the LIDAR PC process ({\bf adq} and {\bf alignment}), and to handle signals to the stepper motor drivers. Therefore, the firmware of the Rabbit microcontroller is a “dummy terminal” that only receives message and drives the control signal to the selected motor/relay. After that, it sends an acknowledge message back to the alignment process.

\subsection{Telescope auto-alignment procedure}   \label{AutoAlignment_section}

\begin{figure}[h!]
  \centering
  \includegraphics[width=0.4\textwidth]{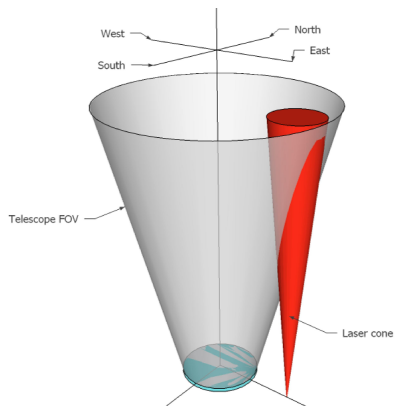}
  \caption{A simplified tree-dimensional model for a bi-axial LIDAR (not to scale), and the reference axis names for futures references.}
  \label{esquemaCoordenadas_fig}
 \end{figure}

The auto-alignment system procedure being applied in this work is based on methods reported in \cite{bib:fiorani} \cite{bib:boliu}, where the overlap between the emission beam and the receiving telescopes  is quantified by averaging the LIDAR signal level at certain range. In this work, the procedure is slightly different to the cited papers, due to the specific features of the multiangle LIDAR being developed.
This algorithm sweeps each FOV's mirror over 2 orthogonal axes (see Figure \ref{esquemaCoordenadas_fig} for a layout of the coordinate system used).

\begin{figure}[h!]
  \centering
  \includegraphics[width=0.4\textwidth]{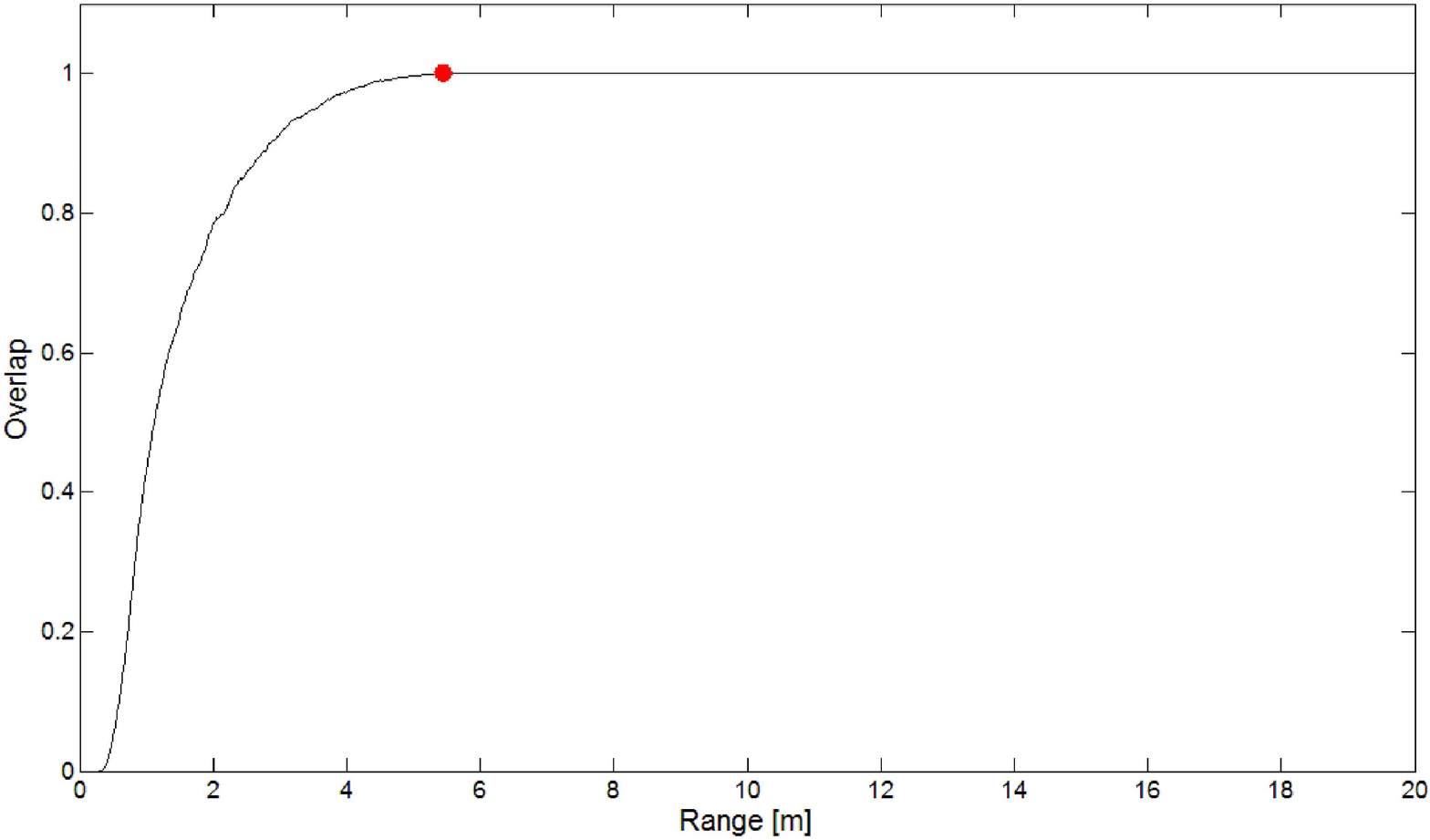}
  \caption{Simulation of the overlap factor with parameters listed in Table \ref{overlap_param_fig} and parallel mirror-laser axes.}
  \label{overlap_fig}
 \end{figure}

An overlap function simulation was done, following the system specifications listed in Table \ref{overlap_param_fig}, and using a laser beam parallel to the reference axis of Figure \ref{overlap_fig}. It can be seen that a full overlap is achieved at 5.445 km approximately, without losses at higher ranges.

\begin{table}[h!]
\begin{center}
\begin{tabular}{|l|c|c|}
\hline Parameter & Value                 \\ \hline
Mirror Diameter    & 40 cm               \\ \hline
Mirror FOV            & 1 mrad            \\ \hline
Distance Mirror-Laser      & 50 cm    \\ \hline
Laser Initial Diameter & 6 mm          \\ \hline
\end{tabular}
\caption{LIDAR features used for the overlap factor calculation.}
\label{overlap_param_fig}
\end{center}
\end{table}

This auto-alignment process is launched when the alignment mode of the {\bf adq} process is set, and the file just acquired is sent to {\bf alignment}. This algorithm evaluates and quantifies the alignment state and saves it with its tilt position. Then, the microcontroller tilts the telescope to a new position and sends an acknowledge message to trigger a new acquisition. When the scanning alignment is finished, the telescope is positioned in the best recorded scenario. 

To find the best alignment angle between the telescope and the laser beam, we record the LIDAR signal level and its angle while scanning through two orthogonal directions (North-South and East-West direction). This cycle is repeated twice, to assure the best overlap scenario possible.

\subsection{Telescope auto-alignment simulation}

\begin{figure}[h!]
  \centering
  \includegraphics[width=0.4\textwidth]{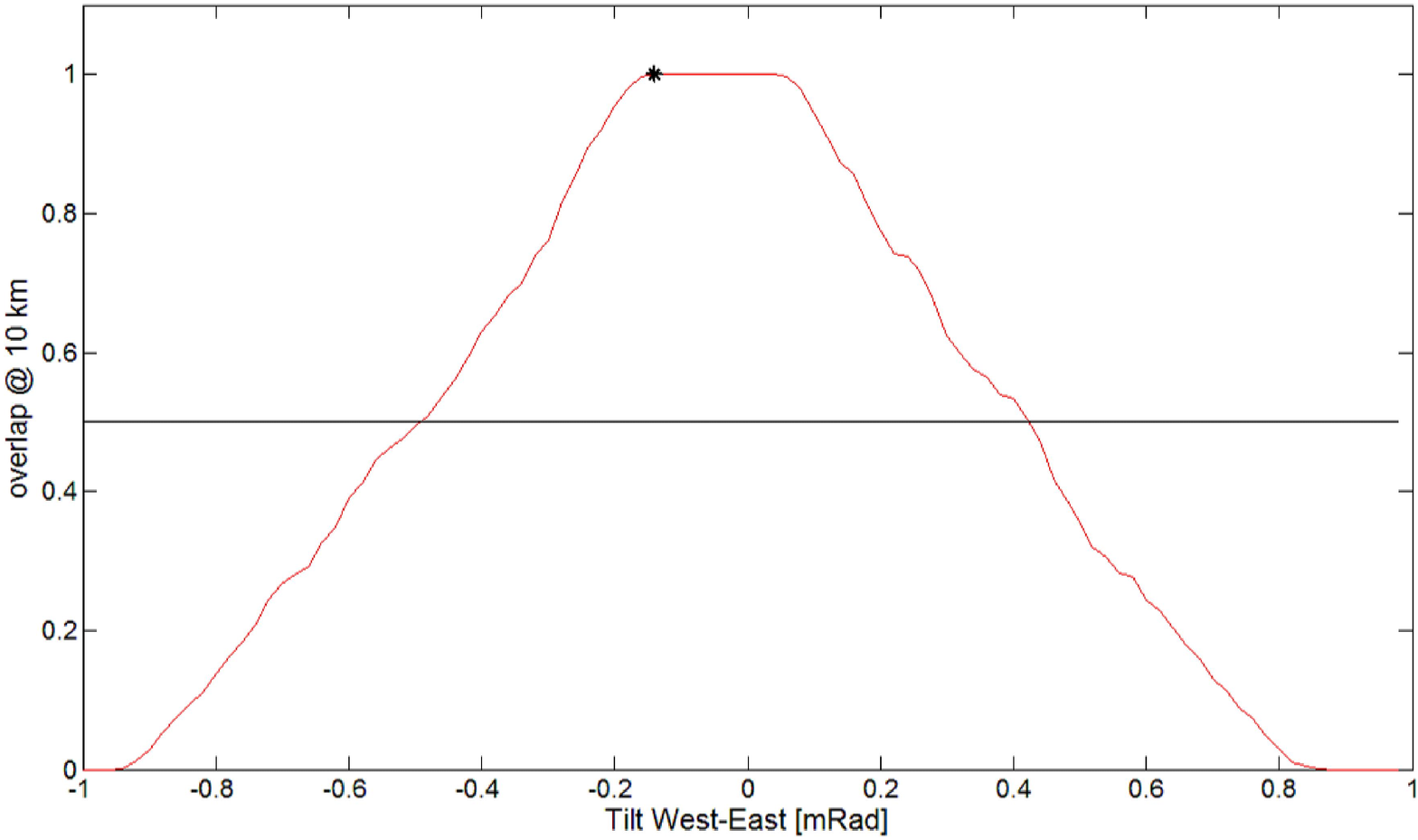}
  \caption{Simulated overlap value at 10 km as a function of East-West axis. The black dot represent an inclination angle of 0.14 mrad.}
  \label{scanWeastEast_fig}
 \end{figure}

\begin{figure}[h!]
  \centering
  \includegraphics[width=0.4\textwidth]{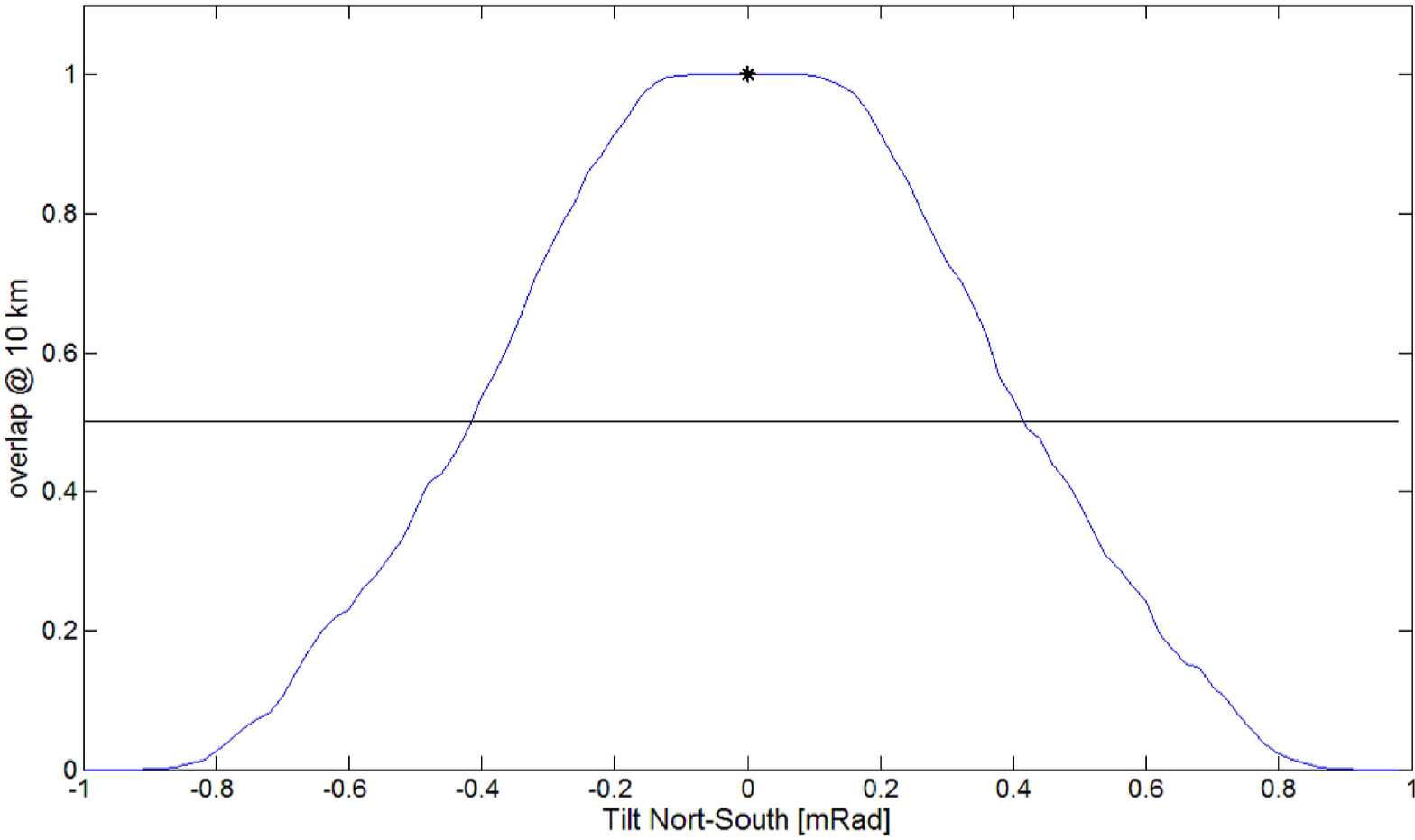}
  \caption{Simulated overlap value at 10 km as a function of North-South axis.}
  \label{scanNorthSouth_fig}
 \end{figure}

Simulation of the alignment process was made over North-South and West-East axis, to show what results from the self-alignment process. The next plots shows different overlap values at 10 km, as a function of scan angle. This altitude has taken based to guarantee good quality signal over most of the troposphere.
Taking into account that the laser beam is homogeneous, and the static atmospheric conditions, we can obtain a trapezoidal shaped plot, as shown in Figures \ref{scanWeastEast_fig} and \ref{scanNorthSouth_fig}, for West-East and North-South span respectively.

For the East-West scanning, the telescope must be set to the angle corresponding to west-side point (-0.14 mrad) of the platform in the trapezoidal plot. This point will assure a full overlap at 10 km, and a extended useful range at lower altitudes.

For scanning over North-South direction (Figure \ref{scanNorthSouth_fig}), the telescope must be set to the middle point of the platform in the trapezoidal form. As can be seen, a symmetric scan is obtained due to the symmetry over West-East axis. This kind of plot allows not only to center the beam inside of the FOV of the telescope, but also to measure the laser divergence and the telescope FOV. From last two figures, it can be readily demonstrated that the width of the rise (or fall) of the trapezoidal plot is equal to the laser divergence, and the full width at half-maximum (FWHM) is the telescope FOV.  

\section{Conclusions}

The construction of the presented multiwavelength scanning Raman LIDAR will be able to provide spectrally-resolved aerosol extinction profiles to characterize the atmospheric transmission at any required line of sight and in a short period of time. The need for a fully automated system was reported, and this is a work-in-progress. The communication and the acquisition systems of the LIDAR are fully operative. First measurements, made at fixed zenith angle, indicate that it is possible to achieve the expected auto-optimization goals during the scanning procedure.
The modularity of the telescope system will permit system maintenance and optimization during operation reducing non-operational times. The collaboration of CEILAP, IFAE/UAB and LUPM to improve their LIDAR systems will permit to attain the requested goals in terms of system construction, LIDAR testing, instrumentation control and LIDAR signal processing. 
Actually, a new azimuth-zenithal scanning bench is under development by Mechanical Department of CITEDEF.

\vspace*{0.5cm}
\footnotesize{{\bf Acknowledgment: }{Authors wish to thank CITEDEF main workshop’s technicians, Mario Proyetti and José Luis Luque from the CEILAP workshop for their support on this development. We gratefully acknowledge support from the following agencies and organizations: Ministerio de Ciencia, Tecnolog\'ia e Innovaci\'on Productiva (MinCyT), Comisi\'on Nacional de Energ\'ia At\'omica (CNEA) and Consejo Nacional  de Investigaciones Cient\'ificas y T\'ecnicas (CONICET) Argentina; State Committee of Science of Armenia; Ministry for Research, CNRS-INSU and CNRS-IN2P3, Irfu-CEA, ANR, France; Max Planck Society, BMBF, DESY, Helmholtz Association, Germany; MIUR, Italy; Netherlands Research School for Astronomy (NOVA), Netherlands Organization for Scientific Research (NWO); Ministry of Science and Higher Education and the National Centre for Research and Development, Poland; MICINN support through the National R+D+I, CDTI funding plans and the CPAN and MultiDark Consolider-Ingenio 2010 programme, Spain; Swedish Research Council, Royal Swedish Academy of Sciences financed, Sweden; Swiss National Science
Foundation (SNSF), Switzerland; Leverhulme Trust, Royal Society, Science and Technologies Facilities Council, Durham University, UK; National Science Foundation, Department of Energy, Argonne National Laboratory, University of California, University of Chicago, Iowa State University, Institute for Nuclear and Particle Astrophysics (INPAC-MRPI program), Washington University McDonnell 
Center for the Space Sciences, USA. The research leading to these results has received funding from the European Union's Seventh Framework Programme ([FP7/2007-2013] [FP7/2007-2011]) under grant agreement nÂ° 262053.}}

\end{document}